\documentclass[runningheads]{llncs}
\usepackage[T1]{fontenc}
\usepackage{graphicx}
\usepackage{booktabs}
\usepackage{comment}
\usepackage{amsmath}
\usepackage{amssymb}
\usepackage[backend=bibtex]{biblatex}
\usepackage[hidelinks]{hyperref}

\begin{document}
\title{Diversified recommendations of cultural activities with personalized determinantal point processes}
\titlerunning{Diversified recommendations of cultural activities with personalized DPPs}
%
\author{Carole Ibrahim\inst{1}\orcidID{0009-0006-2148-5600} \and
Hiba Bederina\inst{2}\orcidID{0009-0003-9926-4099} \and
Daniel Cuesta\inst{1}\orcidID{0009-0000-3445-990X} \and
Laurent Montier\inst{1}\orcidID{0009-0006-7294-8441} \and
Cyrille Delabre\inst{1}\orcidID{0009-0000-9739-5739} \and
Jill-Jênn Vie\inst{2}\orcidID{0000-0002-9304-2220}}
\authorrunning{Ibrahim et al.}
%
\institute{pass Culture \and
Soda Team, Inria Saclay}
\maketitle              
\begin{abstract}
While optimizing recommendation systems for user engagement is a well-established practice, effectively diversifying recommendations without negatively impacting core business metrics remains a significant industry challenge. In line with our initiative to broaden our audience's cultural practices, this study investigates using personalized Determinantal Point Processes (DPPs) to sample diverse \textit{and} relevant recommendations. We rely on a well-known quality-diversity decomposition of the similarity kernel to give more weight to user preferences.
In this paper, we present our implementations of the personalized DPP sampling, evaluate the trade-offs between relevance and diversity through both offline and online metrics, and give insights for practitioners on their use in a production environment.
For the sake of reproducibility, we release the full code for our platform and experiments on GitHub. 

\end{abstract}
\section{Introduction}

The \textit{pass Culture} is a French government initiative launched in 2019 to promote cultural engagement among 3 million young individuals aged 15 to 20. Each user is awarded a fixed credit to spend on the application and access to a diverse array of millions of possible cultural activities (books, cinema, opera, etc.). A central objective of the program is to increase youth participation in cultural activities and to broaden their cultural horizons.
While the Pass Culture has been widely adopted (84\% of 18-year-olds registered by August 2024), a significant portion of the reservations are concentrated on books (especially manga initially, though its share has decreased, with literature now at 40\% of book purchases) and cinema. Other forms of live performance (theater, dance, circus arts) and museums struggle to attract new audiences, accounting for a marginal share of reservations.

A study \cite{llamas2022} suggests that the Pass Culture has only partially succeeded in diversifying cultural practices. While some users report discovering new activities, a significant portion of reservations are for cultural activities they were already familiar with. There's a risk of the Pass primarily intensifying existing cultural practices rather than fostering new ones, particularly among young people who already have higher cultural capital. There is a recognized need to improve the "discoverability" of offers beyond books and recorded music, potentially through enhanced recommendation systems within the application.






To achieve this goal, we investigated the use of Determinantal Point Processes (DPPs) in the recommendation system to enhance diversity in users' recommendations while preserving their engagement. Our results show that DPPs can achieve diversity–utility trade-offs, as described in \cite{wilhelm2018practical}.

In this paper, we describe our approach to experimenting with a DPP-based diversity filter in a production-scale recommender system.
We compare various configurations for incorporating user preferences into diversity sampling.
In particular, our approach is stochastic: if the user refreshes the page, they get different recommendations.
We evaluate the proposed methods in both offline and online settings.

The whole \textit{pass Culture} app code is available under Mozilla Public License on GitHub\footnote{\url{https://github.com/pass-culture}}. Due to the complexity of our production environment, we provide a simplified demonstration of our implementation, available in this GitHub repository\footnote{\url{https://github.com/pass-culture/dpp-paper-demo}}.

\section{Key performance indicators}
\label{kpi}
The universe of all items in the catalog is denoted by $\Omega = \{1, \ldots, N\}$. In our database, $N$ is in the order of millions. 

For each item $i$, we denote its semantic embedding $\phi_i \in \mathbb{R}^{d}$, and we assume that we can compute a quality score $q_i \in [0,1]$ corresponding to the probability of a certain user interacting with this item. This score may rely on a machine learning model trained on historical user interactions.


\paragraph{Relevance} The foremost goal of a recommender system is to provide relevant recommendations. The \emph{relevance} for a set $S \subset \Omega$ of recommended items is $\frac{\sum_{i \in S} q_i}{|S|}$. In online evaluation, it can correspond to a click-through rate.
A model can be trained to predict $q_i$, to rank items by decreasing predicted relevance.

\paragraph{Quality-diversity trade-off}\label{diversity}
We now describe how we model the quality diversity of a set $S$. We note $K$ the similarity kernel between items. In this paper, we focus on the linear kernel, as it simplifies the computation of decomposition: $K_{ij} \propto L_{ij} = q_i q_j \phi_i^T \phi_j.$
%
The \textit{volume} of a set of vectors $(\phi_i)_{i \in S}$ is $Vol(S) = \det((\phi_i^T \phi_j)_{i, j \in S})^{1/2}$. If we consider the kernel above, then
\begin{equation}
\label{eq:dpp_kernel}
\log \det K(S, S) = \sum_{i \in S} \log q_i + 2 \log Vol(S).
\end{equation}
\noindent
This decomposition illustrates the relevance-diversity trade-off of a set $S$ of recommendations, which is of key importance in this paper.

\paragraph{Diversification} 
Moreover, we introduce a business-oriented diversity metric, crucial for communicating usage statistics to our stakeholders, such as the French Ministry of Culture. The metric is defined as follows: each recommendation is assigned a score based on its novelty relative to the user's historical interactions, with a maximum score of 6.5 (2.5 points for a new category, 2 points for a new venue type, 1 point for a new subcategory, and 0.5 points for a new venue or genre). It quantifies the user's increase in coverage of the catalog~\cite{coppolillo2024relevance} and is similar to the notion of serendipity encountered in some works~\parencite{zhang2012auralist}.

\section{Related work}

Determinantal point processes are repulsive point processes that allow sampling points in vector spaces proportionally to the square of the volume formed by their embeddings \parencite{kulesza2012determinantal}, i.e., the determinant of their similarity kernel. They have successfully been used in retrieval and recommendations \parencite{wilhelm2018practical}. The original methods for sampling a subset among $N$ items may have a complexity of $O(N^3)$ due to a kernel decomposition, but it can be reduced to $O(Nd^2)$ in the case of a linear kernel and embeddings of size $d$, and recent techniques have been proposed to cut the complexity to $O(\log N)$ after a preprocessing step of $O(Nd^2)$ \parencite{gillenwater2019tree} or $O(\alpha N)$ where $\alpha < 1$ \parencite{calandriello2020sampling}, which means, without looking at all items.


The closest setting to our approach is \parencite{wilhelm2018practical}. They claim they sample according to a DPP, but in practice, they take a fully \textit{deterministic} approach that consists of computing the greedy max determinant by adding one item at a time. In particular, it means that for a new user, the first recommended set will always be the same, which can harm the diversity of collected data. We decided to adopt the original implementation of DPP, which is \textit{stochastic}, as we assume this may have a positive impact on exploration. \parencite{kathuria2016batched} also suggests that, thanks to its randomization, DPP sampling may achieve lower regret than the greedy max determinant because it allows more exploration.


\section{Recommendation System Overview} \label{sec:system_overview}

\begin{figure*}[h]
    \centering
    \includegraphics[width=\linewidth]{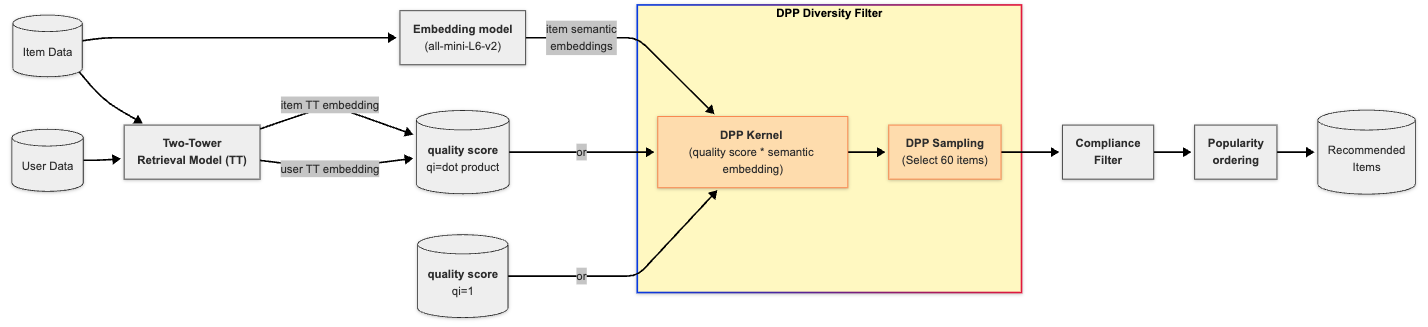}
    \caption{Overview of the recommender architecture with DPP integration.}
    \label{fig:pipeline}
\end{figure*}

We operate a real-time personalized recommendation engine for the pass Culture application. The system follows a multi-stage architecture consisting of three main stages as illustrated in Figure~\ref{fig:pipeline}.

\paragraph{Two-Tower Retrieval model}
The retriever is a two-tower model~\parencite{yi2019sampling,yang2020mixed} from TensorFlow Recommenders\footnote{\url{https://www.tensorflow.org/recommenders}} trained on historical user–item interactions, optimizing recall of clicks. The model learns user and item embeddings using neural networks and stores them in a vector database.
The probability of observing a click depends on the similarity between user and item embeddings. 
It retrieves 1,000 items. 

\paragraph{DPP Diversity Filter} \label{sec:diversification_module}
We implemented a Determinantal Point Process (DPP) module as a post-retrieval step, illustrated in orange in Figure~\ref{fig:pipeline} defined as in Equation \ref{eq:dpp_kernel}. We used \verb+DPPy+, an open-source Python library designed to facilitate the experimental use of DPPs in real-world applications~\parencite{Gautier2019}. Among the top 1,000 items retrieved using the two-tower model embeddings, the DPP module selects a subset of 60 diverse items using the \verb+sample_exact_k_dpp+ class from the \verb+FiniteDPP+ class of DPPy~\parencite{Gautier2019}.

We compute item semantic embeddings $\phi'_i \in \mathbb{R}^{384}$ using the all-MiniLM-L6-v2 model\footnote{\url{https://huggingface.co/sentence-transformers/all-MiniLM-L6-v2}}, to embed item title and description. These embeddings are later reduced to $d = 64$ dimensions for latency constraints \footnote{We observed empirically that embeddings $\phi_i$ of size 64 retain sufficient semantic variation to sample diverse items.}. 



\paragraph{Compliance Filtering \& Popularity Ranking} 
The list of recommendations is filtered to remove any items that do not comply with business rules or user constraints. The resulting items are finally ranked by decreasing popularity.

\section{Evaluation}
 We run offline and online evaluations on the three following recommender pipeline versions:
\begin{itemize}
    \item[A] recommender without the DPP filter (baseline);
    \item[B] recommender with a DPP filter using personalized quality scores $q_i$ based on the two-tower model: $q_i$ represents the cosine similarity between user and item two-tower embeddings to encourage user preferences;
    \item[C] recommender with a DPP filter using a constant value of $q_i = 1$ for all items, relying solely on diversity in semantic space to sample items among the top 1,000 retrieved ones.
\end{itemize}

We refer to those versions as models in the offline metrics and groups in the online metrics.
We present up to 60 items to each user on their homepage.
We evaluate the diversity of recommendations $S$ using the volume and the business diversity metric described in Section~\ref{diversity}. To compute the volumes, we use the non-reduced 384-dimensional item semantic embeddings. We evaluate relevance using the mean cosine similarity between recommended items and users in an offline setting and the click rate in an online setting.

\subsection{Offline evaluation}
For the offline evaluation, we sample 10,000 random users from our database, and we predict their recommendations using the three models, without the compliance filtering. The results are presented in Table~\ref{tab:offline_results}.

\begin{table}[h]
\centering
\begin{tabular}{ccrc} \toprule
& Cosine similarity | & Volume ratio & | Business diversity metric  \\ \midrule
Model A & \textbf{0.525} & 1 & 2.759 \\
Model B & 0.399 & $\times$24.7 & \textbf{3.404} \\ 
Model C & 0.381 & $\times$\textbf{28.8} & \textbf{3.482} \\ \bottomrule
\end{tabular}
\caption{Offline results comparing the baseline (A) with DPP-based recommenders (B and C). Volume ratios are relative to Model A.}
\label{tab:offline_results}
\end{table}

 We observe a notable drop in relevance of approximately 24\% and 27\% for versions B and C, respectively, compared to the baseline. However, this decline is accompanied by a substantial gain in diversity. Specifically, the business diversity metric increases by 23\% and 26\%, while the volume is multiplied by at least 24.7 when filtering with DPP. Although the reduction in relevance is notable, the substantial improvement in diversity offers compelling justification for deploying the DPP-based approach in a production setting, particularly given this paper’s primary objective of enhancing diversity.

\subsection{Online evaluation}

We conducted a 10-day large-scale online A/B/C test comparing the baseline recommender (Group A with 137{,}188 users) with two DPP-based variants: personalized quality scores (Group B with 132{,}063 users) and constant quality scores (Group C with 131{,}458 users).  
The experiment was submitted for ethical approval to our institutional review board. Although explicit consent forms were impractical for this scale, all users had previously agreed to data usage for system improvement during account registration. The A/B/C test results are summarized in Table~\ref{tab:online_results}.

\begin{table}[h]
\centering
\begin{tabular}{ccrc} \toprule
& Click rate | & Volume ratio & | Business diversity metric \\ \midrule
Group A & \textbf{0.54\%} & 1  & 3.132 \\
Group B & 0.34\%* & $\times$12 & \textbf{3.512*} \\ 
Group C & 0.29\%* & $\times$\textbf{15.8}  & \textbf{3.590*} \\ \bottomrule
\end{tabular}
\caption{Online A/B/C test results. Values with * denote statistical significance ($p<0.001$) versus Group A with the Mann-Whitney test.}
\label{tab:online_results}
\end{table}

As anticipated from the offline results, the adoption of DPP resulted in a decrease in relevance: the click-through rate (CTR) dropped by approximately 37\% from the baseline to Group B, and by about 46\% from the baseline to Group C. 
Several factors may contribute to this discrepancy. First, CTR measures user behavior and differs fundamentally from mean cosine similarity, which is used as a proxy for relevance in offline evaluations. Secondly, the offline evaluation was conducted on a subset of 10,000 predominantly active users with multiple interactions in the logs, whereas the online evaluation includes the full user base, encompassing both active and less active users.

In contrast, diversity metrics improved significantly. The volume of distinct items recommended increased by a factor of 12 in Group B and 15.8 in Group C. Business diversity scores also saw meaningful gains, rising by $+12.1\%$ ($p < 0.001$) for Group B and $+14.6\%$ ($p < 0.001$) for Group C. These results highlight a trade-off between relevance and diversity, with the latter showing promising improvements under the DPP framework.

A particularly noteworthy finding is the disparity in click-through rate between the two DPP variants. Specifically, the CTR drops by an additional 14\% when removing the personalized relevance scores ($q_i$). In contrast, the business diversity metric increases by only 2\%. 

\section{Conclusion}

Our findings highlight the potential of DPP sampling to effectively promote diversity in recommendations. 
Based on the A/B/C test results, this work positions DPP as a promising mechanism to foster the discovery of new cultural activities, albeit sometimes at the expense of short-term user engagement.
Despite these promising results, the challenge of maintaining relevance alongside diversity remains an open issue. To address this, we plan to adjust the weight of the quality component in Equation 1, in order to mitigate relevance loss while still enhancing diversity.
This could help identify a Pareto frontier that balances relevance and diversity, providing stakeholders with clear trade-off options to support decision-making.

\begin{credits}
\subsubsection{\ackname} This work was accomplished under a research agreement between Inria, the French Ministry of Culture, and the pass Culture SAS.
We thank Clémence Réda for the interesting discussions and remarks on the paper.
The authors have no competing interests to declare that are relevant to the content of this article.


\end{credits}
%
%
%
%

\printbibliography

\end{document}